\documentclass[12pt]{article}
\usepackage{amsmath}
\usepackage{epsfig}
\usepackage{amssymb,amsfonts}
\usepackage{verbatim}
\usepackage{multirow}

\advance\voffset by -2.0cm
\advance\hoffset by -1.25cm
\def\tr{\,{\rm tr}\, }

\def\be{\begin{equation}}
\def\ee{\end{equation}}
\def\ba{\begin{eqnarray}}
\def\ea{\end{eqnarray}}

\newcommand{\E}{{\cal E}}

\newcommand{\eg}{{\it e.g.~}}
\newcommand{\ie}{{\it i.e.~}}

\newcommand{\id}{{\mathbf{1}}}

\newcommand{\nn}{{\nonumber}}

\begin{document}

\vspace*{-1.5cm}
\thispagestyle{empty}
\begin{flushright}
hep-th/0610175
\end{flushright}
\vspace*{2.5cm}

\begin{center}
{\Large 
{\bf Boundary states, matrix factorisations and
\vspace{0.2cm}

 correlation functions
  for the E-models}} 
\vspace{2.0cm}

{\large Christoph A.\ Keller}%
\footnote{{\tt E-mail: kellerc@itp.phys.ethz.ch}} 
{\large and} 
{\large Sebastiano Rossi}%
\footnote{{\tt E-mail: rossise@itp.phys.ethz.ch}}

\vspace*{0.5cm}

Institut f{\"u}r Theoretische Physik, ETH Z{\"u}rich\\
CH-8093 Z{\"u}rich, Switzerland\\
\vspace*{3cm}

{\bf Abstract}

\end{center}

\noindent 
The open string spectra of the B-type D-branes of the $N=2$ E-models are calculated. Using these results we match the boundary states to the matrix factorisations of the corresponding Landau-Ginzburg models. 
The identification allows us to calculate specific terms in the effective brane superpotential of $E_6$ using conformal field theory methods, thereby enabling us to test results recently obtained in this context. 

\newpage
\renewcommand{\theequation}{\arabic{section}.\arabic{equation}}
\section{Introduction}

$N=2$ minimal models with an $ADE$ classification play a central role
in the description of certain Calabi-Yau compactifications. In
particular, they form the building blocks of Gepner
models \cite{Gepner:1987qi}. For this
reason, there has been great interest in the branes of these models 
and their spectrum.

\noindent In the language of abstract conformal field theory, branes
are given by boundary states. They are linear
combinations of Ishibashi states and must satisfy the Cardy
condition.
On the
other hand, these models can also be described as Landau Ginzburg models. In
this case branes
correspond to matrix factorisations of the 
superpotential \cite{Kapustin:2002bi,Brunner:2003dc,Kapustin:2003ga,
  Kapustin:2003rc,Lazaroiu:2003zi,Herbst:2004ax,Hori:2004ja}. 
 An interesting problem is thus to compare these
descriptions by matching boundary states to matrix factorisations.
For the $A$ and $D$ models, this has been done in
\cite{Brunner:2003dc,Kapustin:2003rc} and \cite{Brunner:2005pq},
respectively.

\vskip .2cm
 In this paper, we perform the match for
the $N=2$ $E$ models.  
For these models the complete set of matrix 
factorisations has been known to mathematicians for some time
\cite{Yoshino,Kajiura:2005yu}. On the CFT side, the
boundary states have been constructed in \cite{Behrend:1999bn,
  Naka:2000he, Lerche:2000jb}. We calculate their spectrum and match
the two descriptions.

\noindent We then use the
identification to discuss obstructions to brane deformations. 
The critical loci of the effective superpotential $W_{eff}$
describe the directions in which a given matrix factorisation
can be deformed, and nonvanishing potential terms describe obstructions to
deformations \cite{Lazaroiu:2001nm,Herbst:2004zm}. On the other hand,
$W_{eff}$ is also the generating 
functional of open string topological disk correlators
\cite{Herbst:2004jp}. Using our
identification, we show that certain specific correlators do not
vanish, so that the brane deformation in these directions is 
obstructed. This calculation can then be used to test
results obtained using other approaches \cite{Knapp:2006rd}. 

\vskip .2cm
\par
 This paper is organised as follows: 
In section \ref{s:bas}, we recall the $ADE$ classification
for affine $su(2)$ models and the construction of their boundary
states. For later use we list some basic properties of
$N=2$ minimal models, the exceptional Lie groups $E_n$, and
matrix factorisations. 
In section \ref{s:Emods}, for each model and each choice of
GSO-projection, we first assemble all information on matrix
factorisations and boundary states. We then calculate their spectrum and
match the boundary states to their corresponding matrix factorisations.
In section \ref{s:corr}, we use this identification to calculate
topological correlators to get certain specific terms of the effective
superpotential. We then draw our conclusions in
section \ref{s:conc}.

\section{Basics} \label{s:bas}

\subsection{Matrix factorisations}
The topological part of a $N=2$ minimal model can also be described in
terms of a 
Landau Ginzburg model. The superpotential $W$ is a weighted homogeneous
polynomial in $x_i$. For $E_n$, the 
superpotentials and the charges $q_i$ of the 
variables are given in table~\ref{tEn}. 
Note that for each model there are two different 
superpotentials which correspond to the two choices of
GSO-projections \cite{Kapustin:2003rc}: the two variable potentials give type $0B$
projection,
the three variable potentials type $0A$.

\noindent B-type branes in the Landau Ginzburg description are given 
by square matrices $E,J$ with polynomial entries, and a charge matrix
$R$. $E,J$ satisfy
\be
E\,J=J\,E=W\,\id\ ,
\ee
or equivalently,
\be
Q^{\,2} = W\, \id \qquad \textrm{where }
Q=\left(\begin{array}{cc}0&J\\E&0\end{array}\right) \ . 
\ee
In our conventions $W$ has $U(1)$ charge 2
and $Q$ has charge 1:
\be
e^{i\lambda R}\,Q(e^{i\lambda q_i}x_i)\,e^{-i\lambda R} = e^{i\lambda}
Q(x_i)\ .
\ee
To determine $R$ uniquely, one must in addition fix $\tr R$ (see
\cite{Kajiura:2005yu} for details).

\noindent Define the operator $D$ by
\be
D(\phi) := Q_2\phi - (-1)^{\deg(\phi)}\phi\, Q_1\ ,
\ee
where $\deg(\phi)$ is the natural $\mathbb{Z}_2$-grading of $\phi$:
even for bosons, odd for fermions. 
The topological spectrum between $Q_1$, $Q_2$ is given by morphisms
$\phi(x_i)$ in the cohomology of $D$.
The charge $q$ of $\phi$ is given by
\be
e^{i\lambda R_2}\,\phi(e^{i\lambda q_i}x_i)\, e^{-i\lambda R_1} =
e^{i\lambda q}\,\phi(x_i)\ . \label{U1mf}
\ee

\noindent The
antibrane $\bar Q$ of $Q$ is obtained by interchanging $E$ and $J$. Note that
the even spectrum between two branes is equivalent to the odd spectrum
between brane and antibrane and vice versa.

\begin{table}[t]
\begin{center}
\begin{tabular}{|c|c|c|cc|c|}
\hline
 & $h$ & $\mathcal{E}$ & &GSO & $q_i$\\
\hline
\multirow{2}{*}{$E_6$} & \multirow{2}{*}{12} &
\multirow{2}{*}{1,4,5,7,8,11}& $W =
x^3+y^4$ &(0B)& \multirow{2}{*}{$[x]=\frac{2}{3},\ [y]=\frac{1}{2},\ [z]=1$}\\ 
& & & $W=x^3+y^4+z^2$&(0A)&\\
\hline
\multirow{2}{*}{$E_7$} & \multirow{2}{*}{18} &
\multirow{2}{*}{1,5,7,9,11,13,17}& $W =
x^3+xy^3$ &(0B)& \multirow{2}{*}{$[x]=\frac{2}{3},\ [y]=\frac{4}{9},\
  [z]=1$}\\  
& & & $W=x^3+xy^3+z^2$&(0A)&\\
\hline
\multirow{2}{*}{$E_8$} & \multirow{2}{*}{30} &
\multirow{2}{*}{1,7,11,13,17,19,23,29}& $W =
x^3+y^5$ &(0B)& \multirow{2}{*}{$[x]=\frac{2}{3},\ [y]=\frac{2}{5},\ [z]=1$}\\ 
& & & $W=x^3+y^5+z^2$&(0A)&\\
\hline
\end{tabular}
\end{center}
\caption{Exceptional groups and their superpotential}\label{tEn}
\end{table}


\subsection{The affine $su(2)$ case}
In this subsection we start the CFT description of $N=2$ minimal models. In view of their construction as cosets (see 2.3)  we will first consider $su(2)$ models.
The  ADE classification gives all possible modular invariant partition
functions  obtained from
combinations of $su(2)_k$ characters. Each such partition function
corresponds  to a simply laced Lie algebra $A_n$, $D_n$, or $E_n$.

\noindent Here we are interested only in the exceptional groups $E_n$. Their
Dynkin diagrams and other properties can be found in
tables~\ref{tEn} and \ref{tDynkin}. The corresponding 
partition functions are given by:
$$
\begin{array}{rlc}
Z_{E_6}=& |\chi_0+\chi_6|^2+|\chi_3+\chi_7|^2+|\chi_4+\chi_{10}|^2 
& (k=10) \\
Z_{E_7}=&  |\chi_0+\chi_{16}|^2+|\chi_4+\chi_{12}|^2 +
|\chi_6+\chi_{10}|^2 + |\chi_8|^2 &\multirow{2}{*}{$(k=16)$}  \\ 
&+\ \chi_8(\hat\chi_2+\hat\chi_{14} +
(\chi_2+\chi_{14})\hat\chi_8 &  \\
Z_{E_8}=&  |\chi_0+\chi_{10}+\chi_{18}+\chi_{28}|^2 +
|\chi_6+\chi_{12}+\chi_{16}+\chi_{22}|^2  &(k=28) 
\end{array}
$$
where the $\chi_\lambda$ are $su(2)_k$ characters, and $k$ is related
to the Coxeter number $h$ of $E_n$ by $h=k+2$.
The boundary states of these model have been constructed some time ago
\cite{Behrend:1999bn}: To each node $L$ of the Dynkin diagram there
corresponds a boundary state given by
\be
|L \rangle = \sum_{l+1\in\mathcal{E}} \frac{\psi_L^{(l)}}{\sqrt{S_0^l}}
  |[l]\rangle\rangle\ .
\ee
Here $l+1$ runs over
the Coxeter exponents of 
$E_n$. The $\psi_L^{(l)}$ for each model
are listed in appendix~\ref{spsi}. The modular transformation
matrix is
\be
S_L^l=\sqrt{\frac{2}{h}}\sin\left(\pi\frac{(L+1)(l+1)}{h}\right) \
. \label{Ssu2} 
\ee

\noindent The overlap of two boundary states is then given by
\be 
\langle\langle L_1||q^{(L_0+\bar
  L_0)/2-c/24}||L_2\rangle\rangle =
\sum_{l=0}^{k}\chi_l(\tilde q)\, n_{lL_1}^{\ \ L_2} \ .
\ee
The matrices $(n_i)_{L_1}^{\ L_2}$ are the so-called fused adjacency
matrices \cite{Behrend:1999bn}. They can be obtained recursively by
applying $su(2)_k$ 
fusion rules 
\be
n_{i+1} = n_1n_i-n_{i-1}\ , \quad i\leq k-1,
\ee
where $n_0$ is the
identity matrix and $n_1$ is the adjacency matrix of the Dynkin diagram. 
By construction the 
$n_i$ form an integer valued representation of the fusion algebra, and
explicit calculation shows that it is non-negative as well. The
$||L\rangle\rangle$ thus satisfy the Cardy condition.


\begin{table}[ht]
\begin{center}
\begin{tabular}{cc}
$E_6$
&
\begin{picture}(120,50)
\put (0,10){\line(1,0){120}}
\put (0,10){\circle*{5}}
\put(0,-3){1}
\put (30,10){\circle*{5}}
\put(30,-3){2}
\put (60,10){\circle*{5}}
\put(60,-3){3}
\put (90,10){\circle*{5}}
\put(90,-3){4}
\put (120,10){\circle*{5}}
\put(120,-3){5}
\put(60,10){\line(0,1){30}}
\put (60,40){\circle*{5}}
\put(60,45){6}
\end{picture} \\

$E_7$ &
\begin{picture}(150,50)
\put (0,10){\line(1,0){150}}
\put (0,10){\circle*{5}}
\put(0,-3){1}
\put (30,10){\circle*{5}}
\put(30,-3){2}
\put (60,10){\circle*{5}}
\put(60,-3){3}
\put (90,10){\circle*{5}}
\put(90,-3){4}
\put (120,10){\circle*{5}}
\put(120,-3){5}
\put (150,10){\circle*{5}}
\put(150,-3){6}
\put(90,10){\line(0,1){30}}
\put (90,40){\circle*{5}}
\put(90,45){7}
\end{picture}
\\

$E_8$ &
\begin{picture}(180,50)
\put (0,10){\line(1,0){180}}
\put (0,10){\circle*{5}}
\put(0,-3){1}
\put (30,10){\circle*{5}}
\put(30,-3){2}
\put (60,10){\circle*{5}}
\put(60,-3){3}
\put (90,10){\circle*{5}}
\put(90,-3){4}
\put (120,10){\circle*{5}}
\put(120,-3){5}
\put (150,10){\circle*{5}}
\put(150,-3){6}
\put (180,10){\circle*{5}}
\put(180,-3){7}
\put(120,10){\line(0,1){30}}
\put (120,40){\circle*{5}}
\put(120,45){8}
\end{picture}
\end{tabular}
\end{center}
\caption{Dynkin diagrams of the exceptional groups} \label{tDynkin}
\end{table}


\subsection{The $N=2$ minimal model}
\label{subsec2.3}
We consider now 
$N=2$ minimal models. Their bosonic subalgebra can be described as the
coset
\be
\frac{su(2)_k\,\oplus\,u(1)_4}{u(1)_{2k+4}}\ .
\ee
The representations of the coset are labelled by triples $(l,m,s)$, where
$l=0,\ldots,k$ is twice the spin of $su(2)$, $m\in \mathbb{Z}_{2k+4}$,
and $s\in \mathbb{Z}_4$. The representations must obey $l+m+s=0 \mod 2$ and are
subject to the identification
\be
(l,m,s) \sim (k-l,m+k+2,s+2)\ .
\ee  
The conformal weights and $U(1)$ charges of the highest weight states
are up to integers given by 
\ba
h(l,m,s)&=&\frac{l(l+2)-m^2}{4(k+2)}+\frac{s^2}{8}\ ,\\
q(l,m,s)&=&\frac{s}{2}-\frac{m}{k+2}\ . \label{U1cft}
\ea
In the NS sector ($s$ even), the chiral primaries appear in the
representations $(l,l,0)$. In the R sector ($s$ odd), the R ground states appear in
$(l,l+1,1)$. 

\noindent The characters $\chi_{[l,m,s]}(q)$ transform under the modular
S-transformation as
\be
\chi_{[L,M,S]}(q)=\sum_{[l,m,s]} S_{LMS}^{\,lms}\, \chi_{[l,m,s]}(\tilde
q)\ ,
\ee
where the sum is over distinct equivalence classes. The $S$-matrix is given by
\be 
S_{LMS}^{\,lms}= 
 \frac{1}{\sqrt{2h}}\,S_L^{\,l}
\,e^{\frac{i\pi}{h} mM} e^{-\frac{i\pi}{2} sS}, \label{S}
\ee
where $S_L^{\,l}$ is the $S$-matrix of $su(2)$ (\ref{Ssu2}).
Let 
\be 
Z=\sum_{l,\bar l}A_{l,\bar l}\chi_l \bar\chi_{\bar l}
\ee
be an ADE-modular invariant of
$su(2)$. 
Then we can construct two different $N=2$ modular invariants by
\cite{Gepner:1989gr} 
\be
Z=\sum A_{l,\bar
    l}\chi_{[l,m,s]} \bar\chi_{[\bar l, m, \pm s]}\ . \label{N=2modinv}
\ee
Physically, the choice $s=\bar s$ corresponds to the type 0B
GSO-projection, and $s=-\bar s$ to type 0A.
See \cite{Gannon:1996hp} for the complete list of all possible modular
invariants of $N=2$ 
superconformal minimal models.

\noindent We want to construct boundary states $||B\rangle\rangle$
that satisfy B-type gluing conditions 
\ba
(L_n-\bar L_{-n})||B\rangle\rangle =& 0 \ ,\nn\\
(J_n+\bar J_{-n})||B\rangle\rangle =& 0 \ ,\label{glue}\\ 
(G^\pm_r+i\eta\, \bar G^\pm_{-r})||B\rangle\rangle =&0\ , \nn
\ea
where $\eta = \pm 1$ determines the spin structure.
The boundary states of the E-models are then given by \cite{Naka:2000he}
\be
||L,M,S\rangle\rangle =
K\,\sum_{[l,m,s]}\frac{\psi_L^{(l)}}{\sqrt{S_{000}^{lms}}}\,
e^{\frac{i\pi}{h}Mm} e^{-\frac{i\pi}{2}
  sS}|[l,m,s]\rangle\rangle, \label{bs}
\ee
where $h$ is the Coxeter number of the group
and $\psi_L^{(l)}$ are the coefficients of the corresponding
$su(2)$ model. The overall normalisation $K$
depends on the model and the type of GSO-projection.

\noindent The Ishibashi states $|[l,m,s]\rangle\rangle$
live in sectors with $m=-\bar m$ and $s=-\bar s$, and the sum in
(\ref{bs}) is over distinct equivalence classes. $||L,M,S\rangle\rangle$
satisfies (\ref{glue}) with $\eta =1$ ($\eta=-1$)
for $S$ even ($S$ odd).
 In section~\ref{s:Emods} we will discuss the exact ranges of $l,m,s$
and $L,M,S$ for each case individually. 

\noindent The chiral primaries $(l,l,0)$ in the overlap between
two boundary states $||B_1\rangle\rangle$ and
$||B_2\rangle\rangle$ should then correspond one-to-one to the
morphisms in the cohomology between the two corresponding matrix
factorisations $Q_1$, $Q_2$ ---  in particular, their $U(1)$
charges given by (\ref{U1mf}) and (\ref{U1cft}) respectively, must be equal.
By calculating and comparing the spectra, we can thus match
matrix factorisations to boundary states.


\section{The exceptional models: $E_6,\ E_7,\ E_8$} \label{s:Emods}
\setcounter{equation}{0}

\subsection{Branes of $E_6$}

\subsubsection{Type 0B: $W = x^3+y^4$} \label{s:2v}
This case corresponds to $m=\bar m$, $s=\bar s$ in
(\ref{N=2modinv}). There are 12
Ishibashi states $|[l,m,s]\rangle\rangle$, $l+1\in \E (E_6)$,
$l+m+s$ even, and $m=0$ or
6 depending on the value 
of $l$: 
\be
\begin{tabular}{cccccc}
$|[0,0,0]\rangle\rangle$, & $|[4,0,0]\rangle\rangle$, &
$|[6,0,0]\rangle\rangle$, & $|[10,0,0]\rangle\rangle$, & 
$|[3,6,1]\rangle\rangle$, & $|[7,6,1]\rangle\rangle$, \\  
$|[0,0,2]\rangle\rangle$, & $|[4,0,2]\rangle\rangle$,&
$|[6,0,2]\rangle\rangle$, & $|[10,0,2]\rangle\rangle$, &
$|[3,6,-1]\rangle\rangle$,& $|[7,6,-1]\rangle\rangle$. \label{IshE60B}
\end{tabular}
\ee
The boundary states are given by
\be
||L,M,S\rangle\rangle = \frac{1}{\sqrt 2}\sum
\frac{\psi^{(l)}_L}{\sqrt{S^{lms}_{000}}}\,   
e^{\frac{i\pi}{12}mM} e^{-\frac{i\pi}{2}sS}\, |[l,m,s]\rangle\rangle \
, \label{bse60B}
\ee
where $L=1,\ldots, 6$ and $S,M\in\mathbb{Z}_4$ with $L+M+S$ even, and 
the sum runs over the Ishibashi states (\ref{IshE60B}).

\noindent The map $\tau: S \mapsto S+2$ maps branes to antibranes,
as it changes the sign of the coupling to RR states. Note that 
in this case there is the symmetry
\ba
||2,S\rangle\rangle = \tau(||4,S\rangle\rangle), &
||1,S\rangle\rangle = \tau(||5,S\rangle\rangle), \label{S+2} \\
||3,S\rangle\rangle = \tau(||3,S\rangle\rangle), &
||6,S\rangle\rangle = \tau(||6,S\rangle\rangle)\nn.
\ea
Moreover, we have
$||L,M,S\rangle\rangle = ||L,M+2,S+2\rangle\rangle$. $M$ is thus fixed
by demanding that $L+M+S$ be even, and by (\ref{S+2}) we can restrict
$S$ to 0,1.
This means that we are left with 12 different
boundary states, 6 for each choice of spin structure.  
 Their spectrum is

\begin{multline}
\langle\langle L_1,M_1,S_1||q^{(L_0+\bar
  L_0)/2-c/24}||L_2,M_2,S_2\rangle\rangle = 
\frac{1}{2} \sum_{[l,m,s]}\chi_{[l,m,s]}(\tilde q)
\,\delta^{(2)}(S_1-S_2-s) \\
\times \left( n_{lL_2}^{\ \
    L_1}(1+e^{\frac{i\pi}{2}(S_2-S_1+s+M_2-M_1+m)}) 
+ n_{10-l\,L_2}^{\phantom{10-l\,L_2}L_1}
(1-e^{\frac{i\pi}{2}(S_2-S_1+s+M_2-M_1+m)}) \right)
\ , \label{spe60b}
\end{multline}
where $n_{lL_2}^{\ \ L_1}$ are the fused adjacency matrices for $E_6$.

\noindent There are six matrix factorisation for this model, listed in
appendix \ref{e6mf}. Their spectrum has been calculated in
\cite{Knapp:2006rd}. It agrees with the chiral primary fields of
(\ref{spe60b}) if we make the identifications: 
\be 
Q_L \equiv ||L,M,0\rangle\rangle 
\ee
with $M\in\{0,1\}$ such that $L+M$ even for the spin structure $S=0$, and
\be
Q_L \equiv ||L,M,1\rangle\rangle 
\ee
with $M\in\{1,2\}$ such that $L+M$ odd for $S=1$.

\subsubsection{Type 0A: $W = x^3+y^4+z^2$}\label{se60a}
There are
12 Ishibashi states   
\be
|[l,0,s]\rangle\rangle \qquad l+1\in \mathcal{E} (E_6)\ ,\label{Is60A}
\ee
with $s\in\mathbb{Z}_4$ such that $l+s$ even.
\noindent The boundary states are given by 
\be
||L,S\rangle\rangle =||L,0,S\rangle\rangle = \frac{1}{\sqrt 2}\sum
\frac{\psi^{(l)}_L}{\sqrt{S^{lms}_{000}}}\,   
 e^{-\frac{i\pi}{2}sS}\, |[l,m,s]\rangle\rangle \
, \label{bse60A}
\ee
the sum running over the Ishibashi states (\ref{Is60A}).
We have $L=1,\ldots, 6$ and $S\in\mathbb{Z}_4$, but again the 
symmetry under $\tau$ allows us to restrict $S\in\{0,1\}$, so
that we have 6 boundary states per spin structure.

\noindent Their overlap is 
\begin{multline}
\langle\langle L_1,S_1||q^{(L_0+\bar
  L_0)/2-c/24}||L_2,S_2\rangle\rangle = \\
\sum_{[l,m,s]}\chi_{[l,m,s]}(\tilde q) \left( n_{lL_2}^{\ \
    L_1}\delta^{(4)}(S_1-S_2-s) + 
n_{10-l\,L_2}^{\phantom{10-l\,L_2}L_1}\delta^{(4)}(S_1-S_2+2-s)\right)
\ . \label{spe60a}
\end{multline}

\noindent The matrix factorisations of $W=x^3+y^4+z^2$ are listed in
appendix \ref{e6mf}, and
their spectrum has been calculated in \cite{Kajiura:2005yu} (beware of the
difference in labelling!) It agrees with (\ref{spe60a}) if we identify
\be 
Q_L \equiv ||L,S\rangle\rangle\ .
\ee

\subsection{Switching between GSO-projections}

\ref{s:2v} and \ref{se60a} illustrate nicely how one can change between one
GSO-Projection and the other: One constructs the new branes out of the old
branes by orbifolding by $\tau$. For instance, if we start out
with the type 0A theory, we take the
orbits of all branes that are not invariant,
\ba
||3,M,S\rangle\rangle
=\frac{1}{\sqrt{2}}(||2,S\rangle\rangle+ ||4,S\rangle\rangle)\, \nn\\
||6,M,S\rangle\rangle
=\frac{1}{\sqrt{2}}(||1,S\rangle\rangle+ ||5,S\rangle\rangle)\ . \nn
\ea
We have thus projected out the Ramond part of these branes. 

\noindent On the other hand, a fixed point $||B\rangle\rangle$ of $\tau$ 
corresponds to
a fractional brane which 
must be resolved by adding linear combinations of the new Ramond
Ishibashi states, \ie
\ba
||B_1\rangle\rangle = \frac{1}{\sqrt{2}} ||B\rangle\rangle
  + \textrm{linear combination of new states}\nn \\
||B_2\rangle\rangle = \frac{1}{\sqrt{2}} ||B\rangle\rangle
- \textrm{linear combination of new states}\nn 
\ea  
It can be checked that by this procedure we really obtain the boundary
states (\ref{bse60B}) of the type 0B theory.


\subsection{Branes of $E_7$}
\subsubsection{Type 0B: $W = x^3+xy^3$} \label{se70b}
$E_7$ is insofar different from $E_6$ as the two
GSO-projections have a different number of boundary states. For 
type 0B projection, there are 28 Ishibashi states,
\be
|[l,0,s]\rangle\rangle \qquad l+1\in \mathcal{E} (E_7), \ s\in\{0,2\} \ ,
\ee
and
\be
|[l,9,s]\rangle\rangle \qquad l+1\in \mathcal{E} (E_7), \ s\in\{-1,1\} \ .
\ee
The boundary states are
\be
||L,M,S\rangle\rangle = \frac{1}{2}\sum_{\substack{ 
      l+1\in\mathcal{E} ,\ m = 
    0,9 \\  m+s\ \textrm{even}}}
 \frac{\psi^{(l)}_L}{\sqrt{S^{lms}_{000}}}
e^{\frac{i\pi}{18}mM} e^{-\frac{i\pi}{2}sS} |[l,m,s]\rangle\rangle 
\label{bse70b}
\ee
where $L=1,\ldots 7$, $S=0,1,2,3$ with $L+M+S$
even. This time the $\psi^{(l)}_L$ are the coefficients for the affine $E_7$
model given in appendix~\ref{psie7}. Again, $S$~odd and $S$~even give two
different spin structures with 14 boundary states each.

\noindent The overlap is 
\begin{multline}
\langle\langle L_1,M_1,S_1||q^{(L_0+\bar
  L_0)/2-c/24}||L_2,M_2,S_2\rangle\rangle = \\ 
\frac{1}{2} \sum_{[l,m,s]} \chi_{[l,m,s]}(\tilde q)\,
 n_{lL_1}^{\ \ L_2} 
 \delta^{(2)}(S_1-S_2-s) \left(1 +
   e^{\frac{i\pi}{2}(M_2+S_2-M_1-S_1+m+s)} \right)\ , 
\label{spe70b}
\end{multline}
where the $n_{lL_1}^{\ \ L_2}$ are now the fused adjacency matrices
for $E_7$.

\noindent The matrix factorisations are given in
appendix \ref{e7mf}. Their spectrum agrees with (\ref{spe70b}) if we
make the identification
\be
Q_L \equiv ||L,M,0\rangle\rangle \ , \ \bar Q_L \equiv
||L,M,2\rangle\rangle \ , \label{ide70b0}
\ee
with $M\in\{0,1\}$ such that $L+M$ even, and
\be
Q_L \equiv ||L,M,1\rangle\rangle \ , \ \bar Q_L \equiv
||L,M,3\rangle\rangle \ ,\label{ide70b1}
\ee
with $M\in\{1,2\}$ such that $L+M$ odd.

\subsubsection{Type 0A: $W = x^3+xy^3+z^2$} \label{se70a}
In this case we only have 14 Ishibashi states,
\be
|[l,0,s]\rangle\rangle \qquad l+1\in \mathcal{E} (E_7), \ s\in\{0,2\} \ .
\ee
For the type 0B case, the map
$\tau: S\mapsto S+2$ had no fixed points. It is thus straightforward
to construct the boundary states for the 0A projection by
\be
||L,S\rangle\rangle =\frac{1}{\sqrt{2}}(||L,M,S\rangle\rangle +
||L,M,S+2\rangle\rangle)\ . \label{bse70a}
\ee
This gives the required 14 states. We could also have obtained these
boundary states by using (\ref{bs}) with $K=\frac{1}{\sqrt 2}$.

\noindent The overlap is 
\begin{multline}
\langle\langle L_1,S_1||q^{(L_0+\bar
  L_0)/2-c/24}||L_2,S_2\rangle\rangle \\ 
 =\sum_{[l,m,s]}  n_{lL_1}^{\ \ L_2}\left(\delta^{(4)}(S_1-S_2-s) +
\delta^{(4)}(S_1-S_2+2-s)\right)
\chi_{[l,m,s]}(\tilde q)\ .
\label{spe70a}
\end{multline}
The identification with the matrix factorisations of appendix
\ref{e7mf} is
\be
\hat Q_L \equiv ||L,S\rangle\rangle \ .\label{mfe70a}
\ee

\subsection{Branes of $E_8$}
\subsubsection{Type 0B: $W = x^3+y^5$} \label{se80b}
The $E_8$ model is completely analogous to the $E_7$ model. For the 0B
projection there are 32 Ishibashi states
\begin{align*}
&|[l,0,s]\rangle\rangle \qquad l+1\in \mathcal{E} (E_8), \ s\in\{0,2\} \ ,\\
&|[l,9,s]\rangle\rangle \qquad l+1\in \mathcal{E} (E_8), \ s\in\{-1,1\} \ ,
\end{align*}
and 32 boundary states $||L,M,S\rangle\rangle$, $L=1\ldots 8$,
$S=0,1,2,3$, $M=0,1$, $L+M+S$ even, given by (\ref{bs}) with $K=\frac{1}{2}$.
Their spectrum is identical to (\ref{spe70b}) with $n_{lL_1}^{\ \ L_2}$
replaced by the fused adjacency matrices of $E_8$.
\noindent The identification with the matrix
factorisations of appendix \ref{e8mf} is
\be
Q_L \equiv ||L,M,0\rangle\rangle \ , \ \bar Q_L \equiv
||L,M,2\rangle\rangle \ 
\ee
and
\be
 Q_L \equiv ||L,M,1\rangle\rangle \ , \ \bar Q_L \equiv
||L,M,3\rangle\rangle \  ,
\ee
with $M$ as in (\ref{ide70b0}) and (\ref{ide70b1}).

\subsubsection{Type 0A: $W = x^3+y^5+z^2$} \label{se80a}
Again, we only have 16 Ishibashi states. The 16 boundary states are
constructed just as in (\ref{bse70a}), their spectrum is as in
(\ref{spe70a}) and they are identified with the matrix factorisations
of appendix \ref{e8mf} by
\be
\hat Q_L \equiv ||L,S\rangle\rangle \ .\label{mfe80a}
\ee


\section{Correlators and the effective superpotential}\label{s:corr}
\setcounter{equation}{0}
\subsection{Introduction and motivation}

In this chapter we make use of the previous match between boundary states and matrix factorisations to calculate specific correlators of the $E_6$ model. These explicit calculations of correlators are to be viewed as checks for results that were obtained by other methods.

\noindent On the one hand, one can try to determine open-closed topological disk amplitudes for minimal models by solving the consistency conditions that these correlators have to satisfy \cite{Herbst:2004jp}, in particular, the $A_\infty$-relations and the homotopy version of bulk-boundary crossing symmetry. These two conditions give rise to an underdetermined set of equations for the correlators. In \cite{Herbst:2004jp} it was proposed that a generalised Cardy conditon should be imposed.
\noindent For the A-series of minimal models this method yields unique correlation functions that agree with previous results. For the E models, however, it appears to be unapplicable \cite{Knapp:2006rd}, as in these cases the Cardy condition seems incompatible with the other sewing conditions. The same incompatibility has been observed for the torus \cite{Herbst:2006nn}.

\noindent A second approach, the Massey product algorithm, was illustrated in \cite{Knapp:2006rd}. Here brane deformations are considered in the context of topological Landau-Ginzburg theories (\cite{Hori:2004ja},\cite{Ashok:2004xq}), i.e as deformations of the superpotential and its factorisations:
\begin{eqnarray}
Q_{def}&=&Q+\sum_{\vec{m}\in \mathbb{N}^{\mathbb{N}}}\alpha_{\vec{m}}u^{\vec{m}}+\sum_{\vec{n}\in \mathbb{N}^{\mathbb{N}}}\tilde{\alpha}_{\vec{n}}(u)s^{\vec{n}}\ ,\\
W_{def}&=&W+\sum_{i}s^i\phi_i\ ,
\end{eqnarray}
where the $\phi_i$ are a basis of the space of bulk chiral primary fields, and $\alpha_{\vec{m}}$, $\tilde{\alpha}_{\vec{n}}$ are boundary fields.\\
$N=2$-worldsheet supersymmetry requires 
\be
Q_{def}^2=W_{def}\ ,\label{q^2=W}
\ee
that is 
\begin{eqnarray}
& &\left\{Q,\alpha_{\vec{m}}\right\}=-\sum_{\vec{m}'+\vec{m}''=\vec{m}}\left\{\alpha_{\vec{m}'},\alpha_{\vec{m}''} \right\} \ ,\label{obscond1}\\
& &\left\{\underbrace{Q+\sum\alpha_{\vec{m}}u^{\vec{m}}}_{=:Q'},\tilde{\alpha}_n(u)\right\}=-\sum_{\vec{n}'+\vec{n}''=\vec{n}}\left\{\tilde{\alpha}_{\vec{n}'}(u),\tilde{\alpha}_{\vec{n}''}(u) \right\}+\sum_{\vec{e}_i}\delta_{\vec{n},\vec{e}_i}\phi_i \ . \label{obscond2}
 \end{eqnarray}
If for some $\vec{m}$ and $\vec{n}$ the r.h.s of (\ref{obscond1}) and (\ref{obscond2}) are non-trivial elements in the cohomology of $Q$ and $Q'$ respectively, then the deformations are obstructed. This means that (\ref{q^2=W}) is satisfied only if $u$ and $s$ satisfy some analytic constraints $p_i(u,s)=0$. For the cases considered in \cite{Knapp:2006rd} these expressions  were found to be integrable, $p_i(u,s)\sim \partial_i W_{eff}$. Since the conditions $p_i(u,s)=0$ are related to the $N=2$-supersymmetry on the worldsheet, they are interpreted as F-term equations \cite{Hori:2004ja} for the deformation parameters, viewed as $N=1$ chiral fields in the low-energy theory. Therefore $W_{eff}$ is interpreted as the space-time effective superpotential and, up to reparametrisations, as the generating function of (symmetrised) open-closed topological correlators \cite{Herbst:2004jp}.\\
\noindent Thirdly, \cite{Knapp:2006rd} also proposed a "mixed" approach. In this approach the $A_\infty$-relations are first solved for the bulk insertions set to zero. One then gets rid of the underdetermination of the problem by requiring agreement with the results of the Massey product algorithm. The correlators so obtained are subsequently used as input for solving the problem with bulk deformations. 

\noindent For the factorisation $W=x^3+y^4-z^2=E_1J_1$ of the $E_6$-model, this mixed prescription leads to a uniquely determined solution. After reparametrisation, it corresponds to the effective superpotential calculated using the Massey product algorithm, but only if certain deformation parameters are set to zero. In particular, the superpotentials in this case are \cite{Knapp:2006rd}:\\

{\small
\begin{eqnarray}
W_{eff}^{mixed}(u;s)&=&W_{eff}^{Massey}(u_1, u_4;s_2=0, s_5=0, s_6, s_8=0, s_9, s_{12})\nonumber\\
W_{eff}^{Massey}(u;s)&=&\frac{5}{832}u_1^{13}+\frac{1}{8}u_4u_1^9+\frac{3}{4}u_4^2u_1^5+u_4^3u_1+\frac{1}{352}s_2u_1^{11}\nonumber\\
                 & &+\frac{1}{192}s_2^2u_1^9-\frac{3}{64}s_5u_1^8+\frac{3}{56}s_6u_1^7+\frac{3}{448}s_2^3u_1^7\nonumber\\
		 & &+\frac{1}{16}s_2^2u_4u_1^5-\frac{1}{10}(s_8+\frac{1}{4}s_6s_2)u_1^5-\frac{1}{2}s_5u_4u_1^4\nonumber\\
		 & &+\frac{1}{8}s_9u_1^4-\frac{1}{4}s_2u_4^2u_1^3+\frac{1}{2}s_6u_4u_1^3-\frac{1}{12}(s_8s_2-s_5^2)u_1^3\nonumber\\
		 & &+\frac{1}{4}s_5s_2u_4u_1^2-\frac{1}{4}s_6s_5u_1^2+\frac{1}{4}s_2^2u_4^2u_1-\frac{1}{2}s_2s_6u_4u_1\nonumber\\
		 & &-s_8u_4u_1+(s_{12}+\frac{1}{4}s_6^2)u_1-\underline{\frac{1}{2}s_5u_4^2}+s_9u_4^2+const     \label{Weff}
\end{eqnarray}}
In this section we want to focus on the discrepancy between the two results. In particular, we shall concentrate on the term $\frac{1}{2}s_5u_4^2$ which we have underlined in (\ref{Weff}).  The presence of this term in $W_{eff}^{Massey}$ implies that the corresponding 4-point disk correlator does not vanish. On the other hand, $W_{eff}^{mixed}$ would imply that it vanishes. Note that even though $W_{eff}^{Massey}$ corresponds to the generating function only up to reparametrisation of the deformation parameters, R-charge considerations show that the term we are considering cannot be eliminated by such a reparametrisation.

\noindent Our task is thus to check if the topological disk correlator $\langle\phi_7\psi_4\int [G,\psi_4]\rangle_1$ vanishes, where $\phi_i\ (\psi_i)$ is the bulk (boundary) field of R-charge $i$, and the label 1 indicates that we impose  boundary conditions corresponding to $Q_1$.

\subsection{Decomposition of $E_6$} \label{ssdec}
For later use we make the following observation:
The fact that 
$c_{k=10}=c_{k=1}+c_{k=2}$ 
suggests that we can decompose $E_6$ into the simpler models $A_1$
and $A_2$. In terms of the LG
potential, this corresponds to the 
observation that $W=x^3+y^4$ is the sum of two 
$A$-model potentials. 
\noindent By decomposing $k=10$ characters 
we can identify
\ba
||0,0\rangle\rangle_1 \otimes ||1,0\rangle\rangle_2 \sim
||1,0\rangle\rangle_{E_6}\ , \nn \\
||0,0\rangle\rangle_1 \otimes ||1,2\rangle\rangle_2 \sim
||5,0\rangle\rangle_{E_6}\ , \label{facbs} \\
||0,0\rangle\rangle_1 \otimes ||0,0\rangle\rangle_2 \sim
||6,0\rangle\rangle_{E_6} \nn \ .
\ea
Here $||L,S\rangle\rangle_{1,2}$ are boundary states of the A-model,
see \eg \cite{Brunner:2005pq} for details of the notation.
\noindent The other $E_6$ boundary states cannot be written as tensor
products of $A_1$ and $A_2$ boundary states. 
This is confirmed by looking at the matrix factorisation
of \ref{e6mf}: $Q_1$, $Q_5$, and $Q_6$ are tensor products, all
the other $Q$ contain terms of the form $xy$ and cannot be decomposed.

\subsection{Topological correlators}
To obtain a topological conformal field theory, one can twist a $N=2$
superconformal model. On the sphere, this leads to a $U(1)$ background
charge of $-\frac{c}{3}$. This means that all topological correlators
vanish unless their total charge is equal to $\frac{c}{3}$.
If we want to calculate such correlators in the original $N=2$ theory,
we must introduce by hand additional fields of charge
$\frac{c}{3}$. We will do this by inserting one unit of spectral flow
$\rho(\xi)$ on the boundary. To get a topological correlator, we then
multiply the result by $\xi^{c/3}$ and let $\xi\rightarrow\infty$
\cite{Warner:1993zh}.

\subsection{Calculating $\langle\phi_7\psi_4\int [G,\psi_4]\rangle_1$} \label{ss:s5u42}

\noindent The matrix factorisation $Q_1$ of the three-variable case
factorises as \footnote{Note that here we use $W=x^3+y^4+z^2$, in agreement with our earlier conventions.}
\be
Q_1 = \left( \begin{array}{cc} 0 & x \\ x^2 & 0 \end{array} \right) \odot 
\left( \begin{array}{cc} 0 & y^2-iz \\ y^2+iz & 0 \end{array} \right) \ ,
\ee
where $\odot$ is the graded tensor product \cite{Hori:2004ja}.
For its fermionic spectrum we get
\ba
\psi :=& 
 \left( \begin{array}{cc} 0 & 1 \\
-x & 0 \end{array} \right) \otimes  
\left( \begin{array}{cc} 1 & 0 \\ 0&1 \end{array} \right)\ , 
&
\psi_2 = y\,\psi
\ . \nn
\ea

\noindent
By comparing charges, we find that
\begin{equation}
\begin{array}{ccc}
\phi_7&\longleftrightarrow&xy\\
\psi_4&\longleftrightarrow&\psi\\
\end{array}.
\end{equation} 
Therefore the superpotential term we are interested in corresponds to the correlator 
\be
D = \langle xy\, \psi \int\! dt\, (G^-_{-1/2}\psi)(t) \rangle \ .
\ee
After factorising we obtain
\begin{multline}
\int dt \langle x \, \left( \begin{array}{cc} 0&1\\ -x&0 \end{array}\right) \, 
\left( G^-_{-1/2} \left( \begin{array}{cc} 0&1\\ -x&0 \end{array}
  \right) \right)\!(t)\rangle_{A_1}\,
\langle\, y\, \id\, \id(t)\, \rangle_{A_2}+  \\
\int dt \langle x \, \left( \begin{array}{cc} 0&1\\ -x&0 \end{array}\right) \, 
\left( \begin{array}{cc} 0&1\\ -x&0 \end{array}
  \right)\!(t)\rangle_{A_1}\,
\langle\, y\, \id\, (G^-_{-1/2}\id)(t)\, \rangle_{A_2}\ .
\end{multline}
In the second term, $\langle\cdots \rangle_{A_2}$ vanishes because its
total charge is $\frac{1}{2} - 1=-\frac{1}{2}$ 
instead of the required $\frac{c}{3}=\frac{1}{2}$.
On the other hand, the $A_2$ correlator of the first term is
independent of $t$. As it contains no integrated operator insertions,
we can evaluate it using \cite{Kapustin:2003rc}:
\be
\langle\,y\,\rangle_{A_2}=\frac{1}{2(2\pi i)^2} \oint dy\, dz\,
\frac{y\cdot {\rm  STr} (\partial_y Q\, \partial_z Q)}{\partial_y
  W_{A_2}\partial_z W_{A_2}}
=\frac{i}{4}.
\ee
To evaluate the $A_1$ correlator, we write it as a coset model CFT
correlator. By comparing $U(1)$ charges, we can identify the fields
\ba
x &\longleftrightarrow & \phi_{110}(z)\phi_{110}(\bar z)\ , \nn \\
\left( \begin{array}{cc} 0&1\\ -x&0 \end{array}
  \right) & \longleftrightarrow & \psi_{110}(s) \nn \ .
\ea
Moreover we insert one unit of spectral flow
$\psi_{1-10}(\xi)$. We thus have to calculate the correlator
\be
\int dt \langle \phi_{110}(z)\,\phi_{110}(\bar z)\, \psi_{112}(t)
\psi_{110}(s)\, \psi_{1-10}(\xi)\rangle\ , \label{A1corr}
\ee
where we have used $G^-_{-1/2}\psi_{110} = \psi_{112}$.
Our task is simplified further since the $A_1$ model is
really just the free boson,
\be
\frac{su(2)_1\oplus u(1)_2}{u(1)_3} = u(1)_6\ ,
\ee
and we can identify (see \eg \cite{Maldacena:2001ky})
\ba
\phi_{110} &\longleftrightarrow & e^{\frac{i}{\sqrt{3}}X}\nn \ ,\\
\psi_{112} &\longleftrightarrow & e^{\frac{-i}{\sqrt{3}}2X}\nn\ ,\\
\psi_{1-10} &\longleftrightarrow & e^{\frac{-i}{\sqrt{3}}X} \nn \ .
\ea
Our original boundary state is a B-type brane and corresponds thus to
Neumann boundary conditions for the free boson.
We can use an explicit expression for (\ref{A1corr})
\cite{Polchinski:1998rq}, 
\be
2\pi i\,C\,|z-\bar z|^{1/3}\,|z-s|^{2/3}\,
|z-\xi|^{-2/3}|\xi-s|^{-1/3} \,\int\! dt\, |\xi-t|^{2/3}\,
|s-t|^{-2/3}\,|z-t|^{-4/3}   \ , 
\ee
where $C$ is a regularised functional determinant.
To obtain the topological correlator, we have to multiply by
$|\xi|^{1/3}$ and let $\xi \rightarrow \infty$. 
Exchanging limit and integral, the result is
\be
\langle \cdots \rangle_{A_1}=2\pi i\, C |z-\bar z|^{1/3}|z-s|^{2/3} \,
\int \frac{dt}{|z-t|^{4/3}|s-t|^{2/3}} \neq 0 \ .
\ee
The result of these calculations is thus that $D$ does not vanish. In a similar way, 
one can show that the correlator corresponding to $s_8u_4u_1$
does not vanish either.\\
Our results therefore agree with those obtained with the Massey product algorithm and not with those calculated with the mixed approach.


\section{Conclusion}\label{s:conc}
Our results for the exceptional models conclude the program started
in \cite{Brunner:2003dc,Kapustin:2003rc,Brunner:2005pq}: 
For all $ADE$ models, the match between matrix factorisations and
boundary states is now known. We have also confirmed that the
different GSO-projections correspond to superpotentials with and
without additional $z^2$ terms. 
 
\noindent The identification of matrix factorisations with boundary states
allows one to calculate topological correlators using conformal field
theory methods. In this paper we have demonstrated this for one of the
correlators of the $E_6$ model. While in general this approach is
likely to be complicated, there are cases (for example the correlator
studied in this paper) where this is actually an efficient method. In
any case, it allows one to check terms of the effective superpotential
that characterise the obstructions of matrix factorisations under
deformations. A good general method to determine the effective
superpotential in minimal models is, however, still missing.

\vspace{1cm}

\centerline{\large \bf Acknowledgements}
\vskip .2cm
\noindent This work has been partially supported by the Swiss National
Science Foundation.
We thank Ilka Brunner, Stefan Fredenhagen and
Matthias Gaberdiel for their support and many helpful discussions.


\appendix

\section{Boundary states for $su(2)$} \label{spsi}
\setcounter{equation}{0}
\subsection{Coefficients for $E_6$} \label{psie6}
{\small
$$
\begin{array}{ccccccc}
l= & 0&3&4&6&7&10\\
\psi_1^{(l)} =& (\frac{1}{2}\sqrt{\frac{3-\sqrt{3}}{6}},& \frac{1}{2},&
\frac{1}{2}\sqrt{\frac{3+\sqrt{3}}{6}},&
\frac{1}{2}\sqrt{\frac{3+\sqrt{3}}{6}}, &
\frac{1}{2}, &\frac{1}{2}\sqrt{\frac{3-\sqrt{3}}{6}}) \\

\psi_2^{(l)} =& (\frac{1}{2}\sqrt{\frac{3+\sqrt{3}}{6}},& 
\frac{1}{2},& \frac{1}{2}\sqrt{\frac{3-\sqrt{3}}{6}},&
-\frac{1}{2}\sqrt{\frac{3-\sqrt{3}}{6}},& -\frac{1}{2},& 
-\frac{1}{2}\sqrt{\frac{3+\sqrt{3}}{6}})\\

\psi_3^{(l)} =& (\frac{1}{2}\sqrt{\frac{3+\sqrt{3}}{3}},& 0,&
-\frac{1}{2}\sqrt{\frac{3-\sqrt{3}}{3}},& 
-\frac{1}{2}\sqrt{\frac{3-\sqrt{3}}{3}},& 0,&
\frac{1}{2}\sqrt{\frac{3+\sqrt{3}}{3}})\\

\psi_4^{(l)} =& (\frac{1}{2}\sqrt{\frac{3+\sqrt{3}}{6}},& 
-\frac{1}{2},& \frac{1}{2}\sqrt{\frac{3-\sqrt{3}}{6}},&
-\frac{1}{2}\sqrt{\frac{3-\sqrt{3}}{6}},& \frac{1}{2},& 
-\frac{1}{2}\sqrt{\frac{3+\sqrt{3}}{6}})\\

\psi_5^{(l)} =& (\frac{1}{2}\sqrt{\frac{3-\sqrt{3}}{6}},& -\frac{1}{2},&
\frac{1}{2}\sqrt{\frac{3+\sqrt{3}}{6}},&
\frac{1}{2}\sqrt{\frac{3+\sqrt{3}}{6}}, &
-\frac{1}{2}, &\frac{1}{2}\sqrt{\frac{3-\sqrt{3}}{6}}) \\

\psi_6^{(l)} = & (\frac{1}{2}\sqrt{\frac{3-\sqrt{3}}{3}},&
0,& -\frac{1}{2}\sqrt{\frac{3+\sqrt{3}}{3}},&
\frac{1}{2}\sqrt{\frac{3+\sqrt{3}}{3}},&0,&
-\frac{1}{2}\sqrt{\frac{3-\sqrt{3}}{3}})

\end{array} \label{psi}
$$
}

\subsection{Coefficients for $E_7$}\label{psie7}
{\small
\begin{equation*}
\begin{array}{cccccccc}
l=&0&4&6&8&10&12&16\\
\psi_1^{(l)} =& (a,& c,& b,& \frac{1}{\sqrt{3}},& b,& c,& a)\\
\psi_2^{(l)} =& (e,& f,& d,& 0,& -d,& -f,& -e)\\
\psi_3^{(l)} =& (c,& b,& -a,& -\frac{1}{\sqrt{3}},& -a,& b,& c)\\
\psi_4^{(l)} =& (f,& -d,& -e,& 0,& e,& d,& -f)\\
\psi_5^{(l)} =& (\frac{1}{\sqrt{6}},& -\frac{1}{\sqrt{6}},&
\frac{1}{\sqrt{6}},& 0,& \frac{1}{\sqrt{6}},& -\frac{1}{\sqrt{6}},&
\frac{1}{\sqrt{6}})\\ 
\psi_6^{(l)} =& (d,& -e,& f,& 0,& -f,& e,& -d)\\
\psi_7^{(l)} =& (b,& -a,& -c,& \frac{1}{\sqrt{3}},& -c,& -a,& b)\\
\end{array}
\end{equation*}}
where
{\small
\begin{equation*}
\begin{array}{cccc}
a=&(18+12 \sqrt{3}\cos{\frac{\pi}{18}})^{-\frac{1}{2}},& b=&(18+12
\sqrt{3}\cos{\frac{11\pi}{18}})^{-\frac{1}{2}},\\ 
c=&(18+12 \sqrt{3}\cos{\frac{13\pi}{18}})^{-\frac{1}{2}},& 
d=&(12(1+\cos{\frac{\pi}{9}}))^{-\frac{1}{2}},\\
e=&(12(1+\cos{\frac{5\pi}{9}}))^{-\frac{1}{2}},&
f=&(12(1+\cos{\frac{7\pi}{9}}))^{-\frac{1}{2}}.\\ 
\end{array}
\end{equation*}
}

\subsection{Coefficients for $E_8$}\label{psie8}
{\small
\begin{equation*}
\begin{array}{ccccccccc}
l=&0&6&10&12&16&18&22&28\\
\psi_1^{(l)}=& (a,&f,&c,&d,&d,&c,&f,&a)\\
\psi_2^{(l)}=&(b,&e,&h,&g,&-g,&-h,&-e,&-b)\\
\psi_3^{(l)}=&(c,&d,&-a,&-f,&-f,&-a,&d,&c)\\
\psi_4^{(l)}=&(d,&a,&-f,&-c,&c,&f,&-a,&-d)\\
\psi_5^{(l)}=&(e,&-h,&-g,&b,&b,&-g,&-h,&e)\\
\psi_6^{(l)}=&(f,&-c,&d,&-a,&a,&-d,&c,&-f)\\
\psi_7^{(l)}=&(g,&-b,&e,&-h,&-h,&e,&-b,&g)\\
\psi_8^{(l)}=&(h,&-g,&-b,&e,&-e,&b,&g,&-h)\\
\end{array}
\end{equation*}}
where
{\small
\begin{equation*}
\begin{array}{cccc}
a=&\left[\frac{15(3+\sqrt{5})+\sqrt{15(130+58\sqrt{5})}}{2}\right]^{-1/2},&
b=&\left[15+\sqrt{75-30\sqrt{5}}\right]^{-1/2},\\ 
c=&\left[\frac{15(3+\sqrt{5})-\sqrt{15(130+58\sqrt{5})}}{2}\right]^{-1/2},&
e=&\left[15-\sqrt{75+30\sqrt{5}}\right]^{-1/2},\\ 
d=&\left[\frac{15(3-\sqrt{5})-\sqrt{15(130-58\sqrt{5})}}{2}\right]^{-1/2},&
g=&\left[15+\sqrt{75+30\sqrt{5}}\right]^{-1/2},\\ 
f=&\left[\frac{15(3-\sqrt{5})+\sqrt{15(130-58\sqrt{5})}}{2}\right]^{-1/2},&
h=&\left[15-\sqrt{75-30\sqrt{5}}\right]^{-1/2}\ .\\ 
\end{array}
\end{equation*}}


\pagebreak
\section{Matrix factorisations}\label{smf}
\setcounter{equation}{0}
\subsection{Matrix factorisations for $E_6$}\label{e6mf}
The matrix factorisations for $W = x^3+y^4$ are \cite{Yoshino}
{\small
\begin{equation}
\begin{array}{cccc}
E_1=J_5 =& \left(\begin{array}{cc}
x&y\\
y^3&-x^2
\end{array}\right) &
E_5=J_1 =& \left( \begin{array}{cc}
x^2&y\\
y^3&-x
\end{array}\right)  \\

E_2=J_4 =& \left(\begin{array}{ccc}
x^2&-xy&y^2\\
y^3&x^2&-xy\\
-xy^2&y^3&x^2
\end{array}\right) &
E_4=J_2 =& \left( \begin{array}{ccc}
x&y&0\\
0&x&y\\
y^2&0&x
\end{array}\right)  \\

E_3=&\left( \begin{array}{cccc}
x&y^2&0&0\\
y^2&-x^2&0&0\\
0&-xy&x^2&y^2\\
y&0&y^2&-x
 \end{array} \right) &
J_3 =& \left( \begin{array}{cccc}
x^2&y^2&0&0\\
y^2&-x&0&0\\
0&-y&x&y^2\\
xy&0&y^2&-x^2
 \end{array} \right)
 \nn \\
 
E_6=&\left( \begin{array}{cc}
x&y^2\\
y^2&-x^2 \end{array} \right) & 
J_6 =& \left( \begin{array}{cc}
x^2&y^2\\
y^2&-x \end{array} \right) \\
\end{array}
\end{equation}
}

The matrix factorisations for $W = x^3+y^4+z^2$ are \cite{Kajiura:2005yu}
{\small
\begin{equation}
\begin{array}{cccc}
E_1 = J_5 =& \!\!\!\!\!\!\!\!\!\!\!\!\!\!\!\!\!\!\!\!\!\!\left( \begin{array}{cc} -y^2+iz & x \\
x^2 & y^2 + iz \end{array} \right)&\!\!\!\!\!\!
J_1 = E_5 =&\!\!\!\!\!\!\!\!\!\!\!\!\!\!\!\!\!\!\!\!\!\! \left( \begin{array}{cc} -y^2-iz & x \\
x^2 & y^2 -  iz \end{array} \right)  \\

E_2 = J_4 =&\!\!\!\!\!\!\!\!\!\!\!\!\!\!\! \left( \begin{array}{cccc}
-y^2+iz &0&xy&x\\
-xy&y^2+iz&x^2&0\\
0&x&iz&y\\
x^2&-xy&y^3&iz \end{array}\right) &\!\!\!\!\!\!\!\!\!\!\!\!\!\!\!
E_4 = J_2 =&\!\!\!\!\!\!\!\!\!\! \left( \begin{array}{cccc}
-y^2-iz&0&xy&x\\
-xy&y^2-iz&x^2&0\\
0&x&-iz&y\\
x^2&-xy&y^3&-iz
\end{array} \right)  \\

E_3 =&\!\!\!\!\!\!\!\!\!\!\!\! \left( \begin{array}{cccccc}
-iz&-y^2&xy&0&x^2&0\\
-y^2&-iz&0&0&0&x\\
0&0&-iz&-x&0&y\\
0&xy&-x^2&-iz&y^3&0\\
x&0&0&y&-iz&0\\
0&x^2&y^3&0&xy^2&-iz
\end{array} \right) &\!\!\!\!\!\!\!\!\!\!\!\!
J_3 =&\!\!\!\!\!\!\!\!\!\!\!\!\!\!\!\!\!\! \left( \begin{array}{cccccc}
iz&-y^2&xy&0&x^2&0\\
-y^2&iz&0&0&0&x\\
0&0&iz&-x&0&y\\
0&xy&-x^2&iz&y^3&0\\
x&0&0&y&iz&0\\
0&x^2&y^3&0&xy^2&iz
\end{array} \right) \\

J_6 =  E_6 =&\!\!\!\!\!\!\!\!\!\!\!\!\!\!\!\!\!\!\!\!\!\! \left( \begin{array}{cccc} 
-z &0& x^2 &y^3 \\
0& -z & y & -x \\
x & y^3 &  z & 0 \\
y & -x^2 & 0 &  z \end{array} \right) &&\\
\end{array}\nn
\end{equation}
}

\subsection{Matrix factorisations $E_7$}\label{e7mf}
For $W = x^3+xy^3$, the matrix factorisations are given by \cite{Yoshino}
\nopagebreak
{\small
\begin{equation}
\begin{array}{cccc}
E_1 =& x&   J_1 =& x^2+y^3  \\

E_2 =& \left(\begin{array}{cc}
x^2&y^2\\
xy&-x\end{array}\right)& 
J_2 =& \left(\begin{array}{cc}
x&y^2\\
xy&-x^2 \end{array}\right)  \\

E_3 =& \left(\begin{array}{ccc}
x^2&-y^2&-xy\\
xy&x&-y^2\\
xy^2&xy&x^2 \end{array}\right) &
J_3 =& \left(\begin{array}{ccc}
x&0&y\\
-xy&x^2&0\\
0&-xy&x \end{array}\right)\\ 

E_4 =& \left(\begin{array}{cccc}
x&y&-y&0\\
y^2&-x&0&-y\\
0&0&x^2&xy\\
0&0&xy^2&-x^2 \end{array}\right) &
J_4 =& \left(\begin{array}{cccc}
x^2&xy&y&0\\
xy^2&-x^2&0&y\\
0&0&x&y\\
0&0&y^2&-x \end{array}\right)\\

E_5 =& \left(\begin{array}{ccc}
y&0&x\\
-x&xy&0\\
0&-x&y \end{array}\right) &
J_5 =& \left(\begin{array}{ccc}
xy^2&-x^2&-x^2y\\
xy&y^2&-x^2\\
x^2&xy&xy^2 \end{array}\right)\\

E_6 =&\left(\begin{array}{cc}
x^2&y\\
xy^2&-x \end{array}\right) &
J_6 =&\left(\begin{array}{cc}
x&y\\
xy^2&-x^2 \end{array}\right)\\

E_7 =& \left(\begin{array}{cc}
x^2&xy\\
xy^2&-x^2 \end{array}\right) &
J_7 =& \left(\begin{array}{cc}
x&y\\
y^2&-x \end{array}\right) \\
\\
\end{array}\nn
\end{equation}
}

The other factorisations $\bar Q_i$ correspond to their antibranes and are
given by $\bar E_i = J_i$, $\bar J_i =E_i$.

\noindent For $W=x^3+xy^3+z^2$, the factorisations are constructed
out of the above by
$$
\hat E_i= \hat J_i = \left(\begin{array}{cc}
z\id & J_i\\
E_i & -z\id \end{array}\right) \ , 
$$
so that $\hat Q_i$ is equal to its own antibrane.

\subsection{Matrix factorisations $E_8$}\label{e8mf}
For $W=x^3+y^5$ the matrix factorisations are given by \cite{Yoshino}
{\small
\begin{equation}
\begin{array}{cccc}
E_1 =& \left(\begin{array}{cc}
x^2&y\\
y^4&-x \end{array}\right) &
J_1 =& \left(\begin{array}{cc}
x&y \\
y^4&-x^2 \end{array}\right) \nn \\

E_2 =& \left(\begin{array}{ccc}
y^4&xy^3&x^2\\
-x^2&y^4&xy\\
-xy&-x^2&y^2 \end{array}\right) &
J_2 =& \left(\begin{array}{ccc}
y&-x&0\\
0&y&-x\\
x&0&y^3 \end{array}\right) \nn \\

E_3 =& \left(\begin{array}{cccc}
0&x^2&-y^3&0\\
-x^2&xy&0&-y^3\\
0&-y^2&-x&0\\
y^2&0&y&-x \end{array}\right) &
J_3 =& \left(\begin{array}{cccc}
y&-x&0&y^3\\
x&0&-y^3&0\\
-y^2&0&-x^2&0\\
0&-y^2&-xy&-x^2 \end{array}\right) \nn \\

E_4 =& \left(\begin{array}{ccccc}
y&-x&0&0&0\\
x&0&0&y^2&0\\
-y^2&0&-x^2&0&-y^3\\
0&-y^2&0&x&0\\
0&0&y^2&y&-x \end{array}\right) &
J_4 =& \left(\begin{array}{ccccc}
y^4&x^2&0&-xy^2&0\\
-x^2&xy&0&-y^3&0\\
0&-y^2&-x&0&y^3\\
-xy^2&y^3&0&x^2&0\\
-y^3&0&-y^2&xy&-x^2 \end{array}\right) \nn \\

E_5 =& \left(\begin{array}{cccccc}
y^4&xy^2&x^2&0&0&xy\\
-x^2&y^3&xy&-x&0&0\\
-xy^2&-x^2&y^3&0&-xy&0\\
0&0&0&y&-x&0\\
0&0&0&0&y^2&-x\\
0&0&0&x&0&y^2 \end{array}\right) &
J_5 =& \left(\begin{array}{cccccc}
y&-x&0&0&0-x\\
0&y^2&-x&xy&0&0\\
x&0&y^2&0&xy&0\\
0&0&0&y^4&xy^2&x^2\\
0&0&0&-x^2&y^3&xy\\
0&0&0&-xy^2&-x^2&y^3 \end{array}\right) \nn \\

E_6 =& \left(\begin{array}{cccc}
x^2&y^2&0&xy\\
y^3&-x&-y^2&0\\
0&0&x&y^2\\
0&0&y^3&-x^2 \end{array}\right) &
J_6 =& \left(\begin{array}{cccc}
x&y^2&0&y\\
y^3&-x^2&-xy^2&0\\
0&0&x^2&y^2\\
0&0&y^3&-x \end{array}\right) \nn \\

E_7 =& \left(\begin{array}{cc}
x&y^2\\
y^3&-x^2 \end{array}\right) &
J_7=& \left(\begin{array}{cc}
x^2&y^2\\
y^3&-x \end{array}\right) \nn \\

E_8 =& \left(\begin{array}{ccc}
y^4&xy^2&x^2\\
-x^2&y^3&xy\\
-xy^2&-x^2&y^3 \end{array}\right) & 
J_8 =& \left(\begin{array}{ccc}
y&-x&0\\
0&y^2&-x\\
x&0&y^2 \end{array}\right) \nn
\end{array}
\end{equation}
}
and their respective antibranes.

\noindent The factorisations for $W=x^3+y^5+z^2$ are constructed in
the same way as for $E_7$.

\end{document}